\documentclass[a4paper,11pt]{article}

\usepackage{jheppub} 

\usepackage{amsfonts}
\usepackage{multirow}
\usepackage{mathrsfs}
\usepackage{graphicx}
\usepackage{amsmath}
\usepackage{amssymb}
\usepackage{bm}
\usepackage{bbm}
\usepackage{color}
\usepackage{hyperref}
\usepackage{array}
\usepackage{diagbox}
\usepackage{float}
\usepackage{makecell}
\usepackage{physics}
\usepackage[caption=false]{subfig}
\allowdisplaybreaks[4]

\def\OMIT#1{}

\def\hlinew#1{%
  \noalign{\ifnum0=`}\fi\hrule \@height #1 \futurelet
   \reserved@a\@xhline}
\usepackage{array}
\newcommand{\PreserveBackslash}[1]{\let\temp=\\#1\let\\=\temp}
\newcolumntype{C}[1]{>{\PreserveBackslash\centering}p{#1}}
\newcolumntype{R}[1]{>{\PreserveBackslash\raggedleft}p{#1}}
\newcolumntype{L}[1]{>{\PreserveBackslash\raggedright}p{#1}}

\newcommand{\nn}{\nonumber}

\newcommand{\beq}{\begin{equation}}
\newcommand{\eeq}{\end{equation}}
\newcommand{\bqa}{\begin{eqnarray}}
\newcommand{\eqa}{\end{eqnarray}}

\newcommand\fverb{\setbox\fverbbox=\hbox\bgroup\verb}
\newcommand\fverbdo{\egroup\medskip\noindent%
			\fbox{\unhbox\fverbbox}\ }
\newcommand\fverbit{\egroup\item[\fbox{\unhbox\fverbbox}]}
\newbox\fverbbox

\makeatletter

\newcommand{\Rmnum}[1]{\expandafter\@slowromancap\romannumeral #1@}
\makeatother

\DeclareCaptionFormat{cont}{#1 (cont.)#2#3\par}

\usepackage{cleveref}[2012/02/15]

\crefformat{footnote}{#2\footnotemark[#1]#3}

\begin{document}
\title{\mbox{}\\[10pt]
Producing Fully-Charmed Tetraquarks via Charm Quark Fragmentation in Colliders}


\author[a]{Xiao-Wei Bai,}
\author[b,c]{Feng Feng,}
\author[a]{Chang-Man Gan,}
\author[d,e]{Yingsheng Huang\footnote{\label{ca}Corresponding author. },}
\author[a]{Wen-Long Sang\cref{ca},}
\author[f]{and Hong-Fei Zhang}


\affiliation[a]{School of Physical Science and Technology, Southwest University, Chongqing 400700, China\vspace{0.2cm}}
\affiliation[b]{Institute of High Energy Physics and Theoretical Physics Center for Science Facilities, Chinese Academy of Sciences, Beijing 100049, China\vspace{0.2cm}}
\affiliation[c]{China University of Mining and Technology, Beijing 100083, China\vspace{0.2cm}}
\affiliation[d]{High Energy Physics Division, Argonne National Laboratory, Argonne, IL 60439, USA}
\affiliation[e]{Department of Physics \& Astronomy,
	Northwestern University, Evanston, IL 60208, USA}
\affiliation[f]{College of Big Data Statistics, Guizhou University of Finance and Economics, Guiyang 550025, China\vspace{0.2cm}}

\emailAdd{xiaoweibai22@163.com}
\emailAdd{f.feng@outlook.com}
\emailAdd{cm.gan@outlook.com}
\emailAdd{yingsheng.huang@northwestern.edu}
\emailAdd{wlsang@swu.edu.cn}
\emailAdd{shckm2686@163.com}

\abstract{
Within the framework of nonrelativistic QCD (NRQCD), we calculate the fragmentation function for a charm quark into an $S$-wave 
fully-charmed tetraquark, denoted as $T_{4c}$.  
The charm-to-$T_{4c}$ fragmentation function is expressed as a sum of 
products of the perturbatively calculable short-distance coefficients and the nonperturbative long-distance matrix elements (LDMEs). 
The short-distance coefficients are ascertained through the perturbative matching 
procedure at lowest order in $\alpha_{s}$ expansion.
The LDMEs are approximated using the $T_{4c}$ four-body wave functions at the origin, which have been evaluated by various phenomenological potential models in literature.
Incorporating the celebrated QCD factorization  and the charm-to-$T_{4c}$ fragmentation function,  we predict the $T_{4c}$ production rate
at high transverse momentum $p_T$ regime in colliders. 
Both the differential distribution over $p_T$ and the integrated cross sections are predicted at the \texttt{LHC}. 
The cross sections for $T_{4c}$ states production can reach several femtobarns to several hundreds femtobarns, 
suggesting a substantial potential for $T_{4c}$ event production at the \texttt{LHC}.
Additionally, we estimate for the photoproduction of $T_{4c}$ in electron-proton ($ep$) collisions. 
It is observed that the cross sections for these processes are moderate at the \texttt{HERA} and \texttt{EIC}, and relatively small at the \texttt{EicC}. 
Given the luminosities of these colliders, the prospect of detecting these fully-charmed tetraquarks at $ep$ colliders is somewhat challenging.
}
\maketitle

\section{Introduction}
The \texttt{LHCb} collaboration has recently reported an unexpected discovery of a new resonance at approximately 6.9 GeV in the di-J/$\psi$ invariant mass spectrum,
labeled as $X(6900)$, with a significance greater than 5$\sigma$~\cite{LHCb:2020bwg}. 
This finding was soon corroborated by both the \texttt{ATLAS} and \texttt{CMS} collaborations~\cite{ATLAS:2023bft,CMS:2023owd}.
Additionally, several similar resonances have been observed~\cite{CMS:2023owd,Zhang:2022toq,Xu:2022rnl}.
Although various alternative explanations have been considered, such as the probability of charmonium molecules and hybrid structures, 
the newly discovered particle, $X(6900)$, is a strong candidate for 
the compact fully-charmed tetraquark, which will be referred to as $T_{4c}$ henceforth. 

Long before the discovery of the $X(6900)$, studies on fully-heavy tetraquarks were already explored,  with some dating back 
four decades~\cite{Iwasaki:1976cn,Chao:1980dv,Ader:1981db}.
Since its identification, a multitude of phenomenological models have been 
employed to explore the mass spectra and decay properties of fully-heavy tetraquark states.
These models encompass a wide range of approaches,
including quark potential models~\cite{Wu:2016vtq,Becchi:2020uvq,Debastiani:2017msn,Bedolla:2019zwg,Lu:2020cns,liu:2020eha,Zhao:2020nwy,Zhao:2020jvl,Giron:2020wpx,Ke:2021iyh,Jin:2020jfc,Wang:2022yes,Wu:2024euj,Bai:2016int,Gordillo:2020sgc},
 QCD sum rules~\cite{Chen:2020xwe,Wang:2020ols,Yang:2020wkh,Wan:2020fsk,Zhang:2020xtb}, 
 and Lattice~\cite{Hughes:2017xie}.
 A variety of theoretical studies, including those in Refs.~\cite{Debastiani:2017msn,Bedolla:2019zwg,Lu:2020cns,liu:2020eha,Zhao:2020nwy,Zhao:2020jvl,Wang:2019rdo,Wang:2021kfv,Liu:2021rtn,Yu:2022lak,Mutuk:2021hmi,Tiwari:2021tmz,Faustov:2022mvs} and others, suggest that the $X(6900)$ state 
 is unlikely to be the ground state and can be categorized as either radially or orbitally excited based on the mass spectrum.
While the spectra and decay properties have been studied
at length, research on the production mechanisms of the fully-charmed tetraquark $T_{4c}$ remains relatively limited.
One of the methods applied to this end is the color evaporation model~\cite{Maciula:2020wri,Carvalho:2015nqf}. 
Additionally, duality relations have been used in references~\cite{Karliner:2016zzc,Berezhnoy:2011xy} to predict the 
production cross section of  $T_{4c}$.

In recent years, several groups, within the framework of NRQCD factorization formalism~\cite{Bodwin:1994jh}, have attempted to investigate the $T_{4c}$ production~\cite{Zhang:2020hoh,Feng:2020qee,Zhu:2020xni,Huang:2021vtb,Feng:2020riv,Feng:2023agq,Feng:2023ghc}. 
Feng et al. explicitly constructed the NRQCD operators and the relevant nonperturbative NRQCD matrix elements 
for the $S$-wave $T_{4c}$ production~\cite{Feng:2020qee}.
Ma and Zhang studied the hadroproduction cross sections of $T_{4c}$ and presented the ratio $\sigma (2^{++})/\sigma(0^{++})$ on the 
transverse momentum $p_{T}$  
at the \texttt{LHC}~\cite{Zhang:2020hoh}. 
By employing phenomenological wave functions at the origin to approximate the NRQCD matrix elements, 
Feng et al.  have provided explicit predictions for $T_{4c}$ production rates
at the  \texttt{LHC}~\cite{Feng:2023agq}.
By including small $p_{T}$ resummation, 
Zhu predicted the production of $T_{4c}$ at low $p_{T}$ at the \texttt{LHC}~\cite{ Zhu:2020xni}.  
The exclusive radiative production and inclusive production of $T_{4c}$ at the B factories were also investigated in~\cite{Feng:2020qee,Huang:2021vtb}. 
\footnote{It is worth noting that the electromagnetic and hadronic decays of the $T_{4c}$ were investigated in recent studies~\cite{Sang:2023ncm,Zhang:2023ffe}. }
These studies suggested that hadron colliders, such as the \texttt{LHC}, are more amenable to the detection of the  $T_{4c}$ tetraquark than 
$e^+e^-$ colliders, with a notably greater potential for $T_{4c}$ production. 

Very recently,  the photoproduction of $T_{4c}$ with $1^{+-}$ at electron-proton colliders, exemplified by the \texttt{HERA}, \texttt{EIC}, and \texttt{EicC}, 
was investigated by Ref.~\cite{Feng:2023ghc}. 
At lowest order in $\alpha_s$, the $S$-wave $1^{+-}$ $T_{4c}$ can be produced through the partonic channel
$\gamma g\to T_{4c}g$.
The study reveals that the prospects for observing the photoproduction of $1^{+-}$ $T_{4c}$ at the \texttt{EIC} are promising. 
Conversely, due to $C$-parity conservation, the production of $0^{++}$ and $2^{++}$ tetraquarks are prohibited at some perturbative order.

It is well established that the inclusive production of a hadron at high transverse momentum $p_T$ is predominantly 
governed by the fragmentation mechanism~\cite{Collins:1989gx}.
In~\cite{Feng:2020riv}, the authors computed the fragmentation function for a gluon to $T_{4c}$, and explored the $T_{4c}$ production rate
through gluon fragmentation at the \texttt{LHC}, where the cross section was found to be significant. 
In addition to gluon fragmentation, $T_{4c}$ can also be produced through charm quark fragmentation. 
In this work, we further investigate the fragmentation function for a charm quark to an $S$-wave $T_{4c}$ by 
employing the NRQCD factorization framework. 
Subsequently, we utilize the fragmentation function to forecast $T_{4c}$ production rates at the  \texttt{LHC}, 
\texttt{HERA}, \texttt{EIC}, and \texttt{EicC}. It is worth noting that the $0^{++}$ and $2^{++}$ states can be generated via charm quark fragmentation, 
consequently, in addition to the $1^{+-}$ state, we 
will, for the first time, present the production rates for these two states at the $ep$ colliders. 

The structure of this paper is as follows. In Section \ref{qcd-factorization-formula}, we utilize the QCD factorization theorem to formulate the inclusive production rate of the $T_{4c}$ with high transverse momentum $p_T$ at both the proton-proton ($pp$) and $ep$ colliders.
In Sec.~\ref{def-frag-fun}, we introduce the definition of the fragmentation function for $c\to T_{4c}$. 
In Sec.~\ref{theoretical-framework}, we present the NRQCD factorization formulas for the fragmentation function.  
Additionally, we outline the procedures for determining the short-distance coefficient (SDC).
In Sec.~\ref{SDCs-result}, we introduce the computational tools and present the analytic expressions for various SDCs.
Sec.~\ref{phen-discu} is devoted to the phenomenological analysis and discussion. 
We make a summary in Sec.~\ref{summary}.  

\section{QCD Factorization theorem for high $p_{T}$ production of $T_{4c}$\label{qcd-factorization-formula}}

In a high-energy collision, the high $p_{T}$ production of an identified hadron $H$ is dominated by the fragmentation mechanism. According to QCD factorization theorem~\cite{Collins:1989gx} , the inclusive production rate of the  $T_{4c}$ with large $p_{T}$ in $pp$ collision can be written in the following form
\bqa\label{t4c-cross section}
\mathrm{d} \sigma\left(p p \rightarrow T_{4c}\left(p_{\mathrm{T}}\right)+X\right) &=&\sum_{i,j,a}\int_0^1 \mathrm{d} x_1  \int_0^1 \mathrm{d} x_2 \int_{0}^{1} \mathrm{d} z\; f_{i/p}(x_1,\mu)f_{j/p}(x_2,\mu) \nn\\
&\times & d {\hat \sigma}(i+j \rightarrow a(p_T/z)+X, \mu)  D_{a \rightarrow T_{4c}}\left(z,\mu\right)+{\cal O}(1/p_T),
\eqa
where $f_{i,j/p}$ denote the parton distribution functions (PDFs) of a proton, $d {\hat \sigma}$ represents the
partonic cross section, $\mu$ is the factorization scale, $z\in [0,1]$ is the ratio of the light-cone momentum carried by $T_{4c}$ with respect to the parent parton $a$,
and $D_{a \rightarrow T_{4c}}$ indicates the fragmentation function for parton $a$ into
$T_{4c}$. The gluon fragmentation into $T_{4c}$ was computed in Ref~\cite{Feng:2020riv}. It is the aim of this work to
investigate the charm/anticharm quark fragmentation into $T_{4c}$. The relevant partonic channels are 
$gg\rightarrow c\bar c$, and $q\bar q\rightarrow c \bar c$ with $q$ representing light quarks.  The cross sections for these two channels are ~\cite{ParticleDataGroup:2022pth}
\begin{equation}~\label{eq-partonic-gg}
\begin{aligned}
\frac{d{\hat{\sigma}_{gg\rightarrow c \bar c}}}{d\hat{t}}=
\frac{1}{6} \frac{\pi \alpha_{s}^{2} }{\hat s^{2}}\left ( \frac{\hat{t}}{\hat{u}}+\frac{\hat u}{\hat t}-\frac{9}{4}\frac{\hat t^{2}+\hat u^{2}}{\hat s^{2}} \right ),
\end{aligned}
\end{equation}
and 
\begin{equation}~\label{eq-partonic-qqbar}
\begin{aligned}
\frac{d{\hat{\sigma}_{q \bar q\rightarrow c \bar c}}}{d\hat{t}}=\frac{4}{9} \frac{\pi \alpha_{s}^{2} }{\hat s^{2}}\left ( \frac{\hat t^{2}+\hat u^{2}}{\hat s^{2}} \right ),
\end{aligned}
\end{equation}
respectively,
where $\hat{s}$, $\hat{t}$ and  $\hat{u}$ are the partonic Mandelstam variables.

Similarly, the inclusive photoproduction rate of the  $T_{4c}$ with large $p_{T}$ in $ep$ collision can be expressed as
\bqa\label{t4c-cross section-ep}
\mathrm{d} \sigma\left(e p \rightarrow T_{4c}\left(p_{\mathrm{T}}\right)+X\right) &=&\sum_{i}\int_0^1 \mathrm{d} x_1  \int_0^1 \mathrm{d} x_2 \int_{0}^{1} \mathrm{d} z\; f_{\gamma/e}(x_1)f_{j/p}(x_2,\mu) \nn\\
&\times & d {\hat \sigma}(\gamma+j \rightarrow a(p_T/z)+X, \mu)  D_{a \rightarrow T_{4c}}\left(z,\mu\right)+{\cal O}(1/p_T).
\eqa
The photon flux $f_{\gamma/e}$ represents the electron's PDF to find a photon with definite momentum fraction, and
is determined by the EPA~\cite{Flore:2020jau,Kniehl:1996we}:
\begin{align}
& f_{\gamma / e}\left(x_{1} \right)=\frac{\alpha}{2 \pi} {\left[\frac{1+\left(1-x_{1}\right)^2}{x_{1}} \ln \frac{Q_{\max }^2}{Q_{\min }^2\left(x_{1}\right)}+2 m_e^2 x_{1}\left(\frac{1}{Q_{\max }^2}-\frac{1}{Q_{\min }^2\left(x_{1}\right)}\right)\right]},
\end{align}
where $Q_{\textrm{min}}^2 (x_{1})=m_e^2x_{1}^2/(1-x_{1})$ and $m_e$ is the electron mass.
The value of $Q_{\max }^2$ varies with experiments, whose typical magnitude is around a few $\mathrm{GeV}^2$. 
At lowest order in $\alpha_s$, $T_{4c}$ can be produced through either the partonic process $\gamma q\to gq$, followed by gluon fragmenting
into $T_{4c}$, or the process $\gamma g\to c\bar{c}$, followed by charm quark fragmenting into $T_{4c}$. The cross sections of the relevant partonic channels are  
\begin{equation}
\begin{aligned}
\frac{d{\hat{\sigma}_{\gamma g\rightarrow c \bar c}}}{d\hat{t}}=\frac{2\pi \alpha \alpha_{s} e_{q}^{2}}{\hat s^{2}}
\left (\frac{\hat u}{\hat t}+\frac{\hat t}{\hat u}  \right ),
\end{aligned}
\end{equation}
and 
\begin{equation}
\begin{aligned}
\frac{d{\hat{\sigma}_{\gamma q\rightarrow g  q}}}{d\hat{t}}= \frac{16}{3} \frac{\pi \alpha \alpha_{s} e_{q}^{2}}{\hat s^{2}}(-\frac{\hat u}{\hat s}-\frac{\hat s}{\hat u}).
\end{aligned}
\end{equation}


\section{Definition of charm-to-$T_{4c}$ fragmentation function \label{def-frag-fun}}

In 1982, the gauge-invariant definition of the quark fragmentation function in $d=4-2\epsilon$ dimensions was first given by Collins and Soper~\cite{Collins:1981uw}. For convenience, we adopt light-cone coordinate to work. A d-dimensional vector $W$ in this system can be expressed as 
\begin{subequations}
	\begin{align}
	&W=(W^{+},W^{-},\mathbf{W_{\bot}}),
	\\
	&W^{+}=(W^{0}+W^{d-1})/\sqrt{2},
	\\
	&W^{-}=(W^{0}-W^{d-1})/\sqrt{2}.
	\end{align}
\end{subequations}
The scalar product of two four-vectors $W$ and $V$ is $W\cdot V=W^{+}V^{-}+W^{-}V^{+}-\mathbf{W_\bot \cdot V_\bot}$. For a charm quark to fragment into a hadron $H$, which is defined as 
\bqa \label{t4c:Fragmentation:Function}
& & D_{c \to H}(z,\mu) =
\frac{z^{d-3} }{ 2\pi \times 4 \times N_c }
\int_{-\infty}^{+\infty} \!dx^- \, e^{-i P^+ x^-/z}
\\
&& \times {\rm tr}\left[ n\!\!\!/
\langle 0 | \Psi(0)
\Phi^\dagger(0,0,{\bf 0}_\perp) \sum_{X} |H(P)+X\rangle \langle H(P)+X|
\Phi (0,x^-,{\bf 0}_\perp) \bar{\Psi}(0,x^-,{\bf 0}_\perp) \vert 0 \rangle \right],
\nn
\eqa
where $n^\mu =(0,1,{\bf 0}_\perp)$ indicates a null reference 4-vector and $\Psi$ represents the field of initial quark. $z$ is defined as $z=P\cdot n / k\cdot n=P^{+} / k^{+}$, which denotes the light-cone momentum fraction. $k$ is the momentum of the initial quark.

The eikonal link $\Phi (0,x^-,{\bf 0}_\perp)$ in (\ref{t4c:Fragmentation:Function}), ensuring the gauge invariance of the fragmentation function, is the path-ordered exponential of the gluon field:
\beq\label{Gauge:Link:Definition}
\Phi (0,x^-,{\bf 0}_\perp) = \mathcal{P} \exp
\left[ i g_s \int_{x^-}^\infty d y^- n\cdot A(0^+,y^-,{\bf 0}_\perp) \right],
\eeq
where $\mathcal{P}$ is path ordering.

The amplitude for a charm quark to fragment into the state $H$ can be generated by applying the Feynman rules as outlined in~\cite{Collins:1981uw}. The phase space for a fragmentation function is given by
\bqa
d\Phi_n &=& {4\pi M \over S_n} \delta(k^+-P^+-\sum_{i=1}^n k_i^+) \prod_{i=1}^n \frac{dk^+_i}{2k_i^+}\frac{d^{d-2}k_{i\perp}}{(2\pi)^{d-1}} \theta(k^+_i),\label{phase:space}
\eqa
where $n$ denotes the number of  unobserved particles in the final state,  
$S_{n}$ stands for the statistical factor for identical particles in the final state, $M$ and $P$ refer to the mass and momentum of $H$, respectively, and $k_{i}$ is the momentum of the $i$-th final-state particle.
The factor $2M$ in the phase space originates from the fact that we adopt nonrelativistic normalization for the heavy-quark spinors~\cite{Bodwin:2014bia}. 

The fragmentation function $D_{c\rightarrow T_{4c}}(z, \mu)$ obeys the celebrated
Dokshitzer-Gribov-Lipatov-Altarelli-Parisi (DGLAP) evolution equation:
\beq\label{DGLAP:evolu:eq}
\mu \frac{\partial}{\partial \mu} D_{c \rightarrow T_{4c}}(z, \mu)= \sum_{i\in\{c,g\}} \int_{z}^{1} \frac{\mathrm{d} y}{y} P_{c\rightarrow i}\left(\frac{z}{y}, \mu\right) D_{i \rightarrow T_{4c}}(y, \mu).
\eeq
For concreteness, the quark-to-quark and quark-to-gluon splitting kernels are given at lowest order in $\alpha _{s}$ by~\cite{Feng:2021uct} 
\beq
P_{c\rightarrow c}(z)=C_{F}\left [ \frac{1+z^{2}}{(1-z)_{+}}+\frac{3}{2}\delta (1-z) \right ],
\eeq
and
\beq
P_{c\rightarrow g}(z)=C_{F}\frac{1+(1-z)^{2}}{z}.
\eeq

\section{Factorization for charm-to-$T_{4c}$ fragmentation function \label{theoretical-framework}}
Fragmentation functions for light hadrons are inherently nonperturbative, typically derived through experimental extraction or via nonperturbative computational methods. In contrast, the scenario for the fragmentation functions of fully heavy hadrons differs significantly.
Prior to hadronization, the heavy quark and
antiquark have to be created at a rather short distance $1/m$, 
therefore it is plausible to invoke asymptotic freedom
to factorize the fragmentation
function as the product of the perturbatively calculable SDCs and the nonperturbative
long-distance matrix elements (LDMEs).
To be specific, the fragmentation functions for a charm quark into a fully heavy hadron
$H$ can be written in the following form
\beq
D_{c \rightarrow H}(z)=\sum_{n} d_{n}(z)\left\langle {O}_{n}^{H}\right \rangle,
\label{NRQCD:fac:quarkonium:frag}
\eeq
where $d_n(z)$ are SDCs and $\left\langle {O}_{n}^{H}\right \rangle$ signify various LDMEs. 

Specifically, we primarily focuses on the fully-charmed $S$-wave tetraquarks, carrying the quantum number $0^{++}$, $1^{+-}$ and $2^{++}$.
 Adopting the diquark-antidiquark basis to specify the color onfiguration is convenient. In this context, the color-singlet tetraquark can be decomposed either as ${\bar{\mathbf{3}}\otimes{\mathbf{3}}}$ or ${\mathbf{6}\otimes\bar{\mathbf{6}}}$. 
Due to the Fermi-Dirac statistics,
the former case corresponds to the spin-1 diquark, while the latter corresponds to the spin-0 diquark. According to \eqref{NRQCD:fac:quarkonium:frag}, the fragmentation function for a charm quark into the tetraquark at lowest order in velocity can be written as
\begin{align}
\notag D_{c \rightarrow T_{4 c}}\left(z\right)=
&\frac{d_{3, 3}\left[c \rightarrow c c \bar{c} \bar{c}^{(J)}\right]}{m^{9}}
\left\langle {O}_{3,3}^{(J)}\right \rangle
+\frac{d_{6, 6}\left[c \rightarrow c c \bar{c} \bar{c}^{(J)}\right]}{m^{9}}
\left\langle {O}_{6,6}^{(J)}\right \rangle\\
&+\frac{d_{3, 6}\left[c \rightarrow c c \bar{c} \bar{c}^{(J)}\right]}{m^{9}}
2\mathrm{Re}[\left\langle {O}_{3,6}^{(J)}\right \rangle]+\cdots,
\label{NRQCD:fac:t4c:fragmentation}
\end{align}
where the SDCs $d_n(z)$ are dimensionless, ${O}^{(J)}_{\rm color}$ ($J=0,1,2$) denote the NRQCD production operators, which are defined via
\begin{subequations}
	\begin{align}
	&O^{(J)}_{3,3}=\mathcal{O}^{(J)}_{\bar{\mathbf{3}}\otimes{\mathbf{3}}} 
	\sum_{X}\left. | T^{J}_{4c}+X  \right. \rangle\left.\langle T^{J}_{4c}+X \right.|
	\mathcal{O}^{(J)\dagger }_{\bar{\mathbf{3}}\otimes{\mathbf{3}}},
	\\
	&O^{(J)}_{6,6}=\mathcal{O}^{(J)}_{{\mathbf{6}}\otimes \bar{\mathbf{6}}} 
	\sum_{X}\left. | T^{J}_{4c}+X  \right. \rangle\left.\langle T^{J}_{4c}+X \right.|
	\mathcal{O}^{(J)\dagger }_{{\mathbf{6}}\otimes \bar{\mathbf{6}}},
	\\
	&O^{(J)}_{3,6}=\mathcal{O}^{(J)}_{\bar{\mathbf{3}}\otimes{\mathbf{3}}} 
	\sum_{X}\left. | T^{J}_{4c}+X  \right. \rangle\left.\langle T^{J}_{4c}+X \right.|
	\mathcal{O}^{(J)\dagger }_{{\mathbf{6}}\otimes \bar{\mathbf{6}}}.
	\end{align}
\end{subequations}
$\mathcal {O}^{(J)}_{\rm color}$ represent the composite color-singlet four-quark operators, which can be expressed as~\cite{Huang:2021vtb,Feng:2020riv}
\begin{subequations}
	\begin{align}
	&\mathcal{O}^{(0)}_{\bar{\mathbf{3}}\otimes{\mathbf{3}}}=-\frac{1}{\sqrt{3}}[\psi_a^T(i\sigma^2)\sigma^i\psi_b] [\chi_c^{\dagger}\sigma^i (i\sigma^2)\chi_d^*]\;
	\mathcal{C}^{ab;cd}_{\bar{\mathbf{3}}\otimes{\mathbf{3}}},
	\\
	& \mathcal{O}^{(1)}_{\bar{\mathbf{3}}\otimes{\mathbf{3}}}= -{\frac{i}{\sqrt{2}}} \left[\psi_a^T (i \sigma^2)\sigma^j\psi_b\right]\left[\chi_c^\dagger\sigma^k (i \sigma^2)\chi_d^*\right]\,\epsilon^{ijk}\;{\mathcal C}^{ab;cd}_{\bar{\mathbf{3}}\otimes{\mathbf{3}}},
	\\
	&\mathcal{O}^{(2)}_{\bar{\mathbf{3}}\otimes{\mathbf{3}}}=\frac{1}{2}[\psi_a^T(i\sigma^2)\sigma^m\psi_b] [\chi_c^{\dagger}\sigma^n(i\sigma^2)\chi_d^*]\;\Gamma^{ij;mn}
	\;\mathcal{C}^{ab;cd}_{\bar{\mathbf{3}}\otimes{\mathbf{3}}},
	\\
	&\mathcal{O}^{(0)}_{\mathbf{6}\otimes\bar{\mathbf{6}}}=
	[\psi_a^T(i\sigma^2)\psi_b] [\chi_c^{\dagger}(i\sigma^2)\chi_d^*]\;
	\mathcal{C}^{ab;cd}_{\mathbf{6}\otimes\bar{\mathbf{6}}},
	\end{align}
	\label{NRQCD:composite:operators}
\end{subequations}
where $\sigma^i$ is Pauli matrix, $\psi$ and $\chi^\dagger$ are the standard NRQCD fields annihilating the heavy quark and antiquark, respectively. The symmetric traceless tensor is given by $\Gamma^{ij;mn}\equiv g^{i m} g^{j n}+g^{i n} g^{j m}-\frac{2}{3} g^{i j} g^{mn}$,
and the color projection tensors read
\begin{subequations}
	\bqa
	&& \mathcal{C}^{ab;cd}_{\bar{\mathbf{3}}\otimes{\mathbf{3}}}\equiv \frac{1}{2\sqrt{3}}(\delta^{ac}\delta^{bd}-\delta^{ad}\delta^{bc}),
	\\
	&& \mathcal{C}^{ab;cd}_{{\mathbf{6}}\otimes \bar {\mathbf{6}}}
	\equiv \frac{1}{2\sqrt{6}}(\delta^{ac}\delta^{bd}+\delta^{ad}\delta^{bc}).
	\eqa
	\label{color:tensor}
\end{subequations}

The SDCs in \eqref{NRQCD:fac:t4c:fragmentation} can be determined through the standard perturbative matching procedure. 
Since the SDCs are insensitive to the long-distance dynamics with the 
spirit of factorization,  we can replace the physical $T_{4c}$ state in \eqref{NRQCD:fac:t4c:fragmentation} with a fictitious ``tetraquark" state $\vert [cc][\bar{c}\bar{c}]\rangle$, which composes of a pair of heavy quarks and a pair of heavy antiquarks carrying the same quantum number as the physical $T_{4c}$. It is convenient to label the fictitious state with $\tilde {T}_{4c}$. After this replacement,
we can readily get the desired SDCs by computing both sides of \eqref{NRQCD:fac:t4c:fragmentation} in perturbative QCD and NRQCD. 
For the QCD side, we employ the covariant projector to project out the desired spin/color quantum number of $\tilde {T}_{4c}$
\beq
\bar u^a_i\bar u^b_j v^c_k v^d_l\to (\textsf{C}\Pi_\mu)^{ij}(\Pi_\nu \textsf{C} )^{lk}\mathcal{C}^{ab;cd}_{\rm color} J^{\mu\nu}_{0,1,2},
\eeq
where $\textsf{C}=i\gamma^2\gamma^0$ is the charge conjugation matrix, $i$, $j$, $k$ and $l$ denote the spinor indices, $a$, $b$, $c$ and $d$ are the color indices. $\mathcal{C}^{ab;cd}$ represents
color projection tensor that is given in \eqref{color:tensor}, $\Pi_\mu$ is the spin projector of two fermion~\cite{Petrelli:1997ge}, and the covariant projectors $J^{\mu\nu}_{0,1,2}$ can be expressed as
\begin{subequations}
	\begin{align}
	&J^{\mu \nu }_{0}=\frac{1}{\sqrt{3}}\eta ^{\mu \nu},
	\\
	&J^{\mu \nu }_{1}(\epsilon )=-\frac{i}{\sqrt{2p^{2}}}\epsilon ^{\mu \nu \alpha \beta}\epsilon _{\alpha }p_{\beta },
	\\
	&J^{\mu \nu }_{2}(\epsilon )=\epsilon_
	{\alpha \beta}\left \{ \frac{1}{2}\left [ \eta^{\mu \alpha} \eta^{\nu \beta} +\eta^{\mu \beta} \eta^{\nu \alpha} \right ]
	-\frac{1}{3}\eta^{\mu \nu}\eta^{\alpha \beta}  \right \},
	\end{align}
	\label{C-G:coupling}
\end{subequations}
where $p$ denotes the momentum of $T_{4c}$ and $\eta^{\mu \nu}\equiv -g^{\mu \nu}+p^{\mu}p^{\nu}/p^{2}$. 

For completeness, we present the results for the perturbative LDMEs involving $\tilde {T}_{4c}$ as follows
\begin{equation}
\begin{aligned}
\left\langle {O}_{3,3}^{(J)}\right \rangle=\left\langle {O}_{6,6}^{(J)}\right \rangle=
\mathrm{Re}[\left\langle {O}_{3,6}^{(J)}\right \rangle]=4^{2} (2J+1).
\end{aligned}
\end{equation}

\begin{figure}[hbtp]
	\centering
	\includegraphics[width=1.0\textwidth]{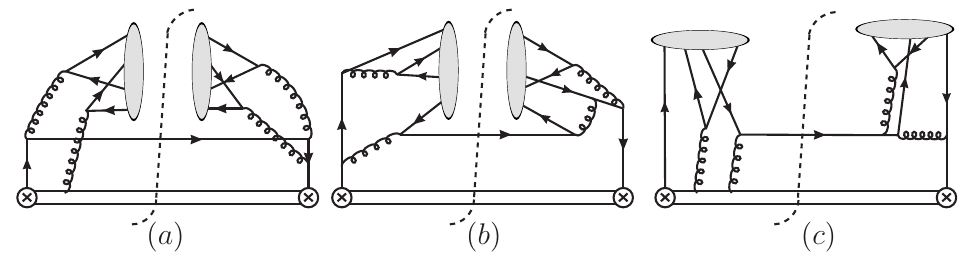}
	\caption{Some representative Feynman diagrams for the fragmentation function of a charm quark into $T_{4c}$, drawn with {\tt JaxoDraw}~\cite{Binosi:2008ig}. The grey blob denotes
		the tetraquark. Horizontal double line indicates the eikonal line.}
	\label{diagrams}
\end{figure}

\section{Computational tools and analytic expressions for the SDCs\label{SDCs-result}}

To expedite the calculation, we employ 
the package {\tt FeynArts}~\cite{Hahn:2000kx} to generate the Feynman diagrams and the corresponding amplitudes for the partonic process $c\to cc\bar{c}\bar{c}$. At tree level, there are 96 diagrams, and some representative Feynman diagrams are depicted in Fig.~\ref{diagrams}.
For practical calculations, it is most convenient to work in the Feynman gauge.
We utilize the packages {\tt FeynCalc/FormLink}~\cite{Mertig:1990an,Feng:2012tk} to handle the Dirac trace algebra and Lorentz contraction. 

The technique outlined in the preceding section is utilized to determine the SDCs.
For a charm quark fragmentation into $0^{++}$ tetraquark, 
the three SDCs in~\eqref{NRQCD:fac:t4c:fragmentation}  are deduced as follows
\begin{equation}
\begin{aligned}
 & d_{6,6}\left(c \rightarrow 0^{++}\right)= \frac{\pi^{2}\alpha_{s}^{4}}{373248 (4-3 z)^6 (z-4)^2 z (11 z-12)(z^2-16 z+16)}  \bigg\{-120 (z-4) \\
&\times  (11z-12) \left(z^2-16 z+16\right) \left(35 z^4-535 z^3+3472 z^2-4240z+512\right) (3 z-4)^5\\
& \times \log \left(z^2-16 z+16\right)-30 (11 z-12) \left(z^2-16 z+16\right) (3395 z^5-48020 z^4+126144 z^3\\
&-75776z^2-38656z+62464) (3 z-4)^5 \log (4-3 z)+75 (11 z-12)\left(z^2-16 z+16\right) (735 z^5\\
&-10684 z^4+34208 z^3-44160z^2 +20224z+9216) (3 z-4)^5 \log\left[\left(4-\frac{11z}{3}\right)(4-z)\right]\\
&+16(z-1)(7916587z^{12}-263987840z^{11}+3125201872 z^{10}-16993694336 z^9+51814689024 z^8\\
&-99638283264 z^7+133459423232 z^6-140136398848 z^5+127161204736z^4-96695746560 z^3\\
&+53372518400 z^2-17930649600z+2717908992)\bigg\},
\end{aligned}
\end{equation}
\begin{equation}
\begin{aligned}
& d_{3, 3}\left(c \rightarrow  0^{++}\right)=\frac{\pi^{2}\alpha_{s}^{4}}{559872 (4-3 z)^6 (z-4)^2 z (11 z-12)\left(z^2-16 z+16\right)} \bigg\{-264 (z-4)\\
&\times (11 z-12) \left(z^2-16 z+16\right) \left(13 z^4-57 z^3-656 z^2+1424z-512\right) (3 z-4)^5\\
&\times \log \left(z^2-16 z+16\right)+6 (11 z-12)\left(z^2-16 z+16\right) (1273 z^5-16764 z^4+11840 z^3\\
&+247808z^2-472320 z+171008) (3 z-4)^5 \log (4-3 z)-3 (11 z-12)\left(z^2-16 z+16\right) (129 z^5\\
&-7172 z^4+49504 z^3-108416z^2+73984 z-9216) (3 z-4)^5 \log\left[\left(4-\frac{11z}{3}\right)(4-z)\right]\\
&+16 (z-1) (657763 z^{12}-10028192z^{11}+188677968 z^{10}-2600899712 z^9+18018056448 z^8\\
&-71685000192z^7+179414380544 z^6-294834651136 z^5+321642168320z^4-229388845056 z^3\\
&+102018056192 z^2-25480396800z+2717908992)\bigg\},
\end{aligned}
\end{equation}
and
\begin{equation}
\begin{aligned}
& d_{3,6}\left(c \rightarrow 0^{++}\right)=\frac{\pi^{2}\alpha_{s}^{4}}{186624 \sqrt{6} (4-3 z)^6 (z-4)^2 z
(11 z-12) \left(z^2-16 z+16\right)} \bigg\{24 (z-4)\\
&\times (11 z-12) \left(z^2-16 z+16\right) \left(225 z^4-3085 z^3+17456 z^2-19760z+1536\right) (3 z-4)^5\\
&\times \log \left(z^2-16 z+16\right)-6 (11 z-12)\left(z^2-16 z+16\right) (555 z^5+52428 z^4-363328 z^3\\
&+616448z^2-270080 z+70656) (3 z-4)^5 \log (4-3 z)-3 (11 z-12)\left(z^2-16 z+16\right) (1245 z^5\\
&-84308 z^4+601696z^3-1333120z^2+914688 z-119808) (3 z-4)^5 \log\left[\left(4-\frac{11z}{3}\right)(4-z)\right]\\      
&+16 (z-1)(1829959z^{12}-44960912 z^{11}+285792656 z^{10}-1090093952z^9+5123084544 z^8\\
&-24390724608 z^7+77450817536 z^6-153897779200z^5+194102034432 z^4-155643543552 z^3\\
&+77091307520 z^2-21705523200z+2717908992)\bigg\}.
\end{aligned}
\end{equation}

The SDC for a charm quark fragmentation into $1^{+-}$ tetraquark is formulated as follows
\begin{equation}
\begin{aligned}
 & d_{3,3}\left(c \rightarrow 1^{+-}\right)=\frac{\pi^{2}\alpha_{s}^{4}}{279936 (4-3 z)^6 (z-4)^2 z (11 z-12)\left(z^2-16 z+16\right)} \bigg\{480 (z-4)\\
 &\times (11 z-12)\left(z^2-16 z+16\right) \left(4 z^4+115 z^3-316 z^2+112z+64\right) (3 z-4)^5\\
 &\times \log \left(z^2-16 z+16\right)+6 (11 z-12)\left(z^2-16 z+16\right) (4825 z^5-56232 z^4+378480z^3\\
 &-942528 z^2+672768 z-60416) (3 z-4)^5 \log (4-3 z)-3 (11z-12) \left(z^2-16 z+16\right) (5465 z^5\\
 &-40392 z^4+254320z^3-722368 z^2+611328 z-101376) (3 z-4)^5 \log\left[\left(4-\frac{11z}{3}\right)(4-z)\right]\\
 &+16 (z-1) z(476423 z^{11}+32559240 z^{10}-934590720 z^9+8015251776z^8\\
 &-35393754624 z^7+94265413632 z^6-160779010048 z^5+177897046016z^4-124600254464 z^3\\
 &+51223461888 z^2-10217324544z+490733568)\bigg\}.
\end{aligned}
\end{equation}

The SDC for a charm quark fragmentation into $2^{++}$ tetraquark reads
\begin{equation}
\begin{aligned}
	 & d_{3,3}\left(c \rightarrow 2^{++}\right)=\frac{\pi^{2}\alpha_{s}^{4}}{2799360 (4-3 z)^6 (z-4)^2 z^2 (11
		z-12) \left(z^2-16 z+16\right)} \bigg\{-33 (11 z-12) \\
		&\times \left(z^2-16z+16\right) \left(3581 z^5+53216 z^4-326176 z^3+419456 z^2-6912z+55296\right) (3 z-4)^6 \\
		&\times \log\left[\left(4-\frac{11z}{3}\right)(4-z)\right]+672 (z-4) (11 z-12) \left(z^2-16 z+16\right) (47
		z^5+12186 z^4\\
		&-44608 z^3+40000 z^2-7936 z+4608) (3 z-4)^5\log \left(z^2-16 z+16\right)+6 (11 z-12)\\
		&\times \left(z^2-16 z+16\right) (107645 z^6-1088988 z^5+7805536 z^4-20734976 z^3+18933504z^2\\
		&-6013952 z+1695744) (3 z-4)^5 \log (4-3 z)+16 (z-1) z (96449507 z^{12}-158520388 z^{11}\\
		&-26228206896z^{10}+281743037888 z^9-1355257362432 z^8+3773988390912z^7\\
		&-6637452959744 z^6+7595797282816 z^5-5643951472640z^4+2662988513280 z^3\\
		&-788934950912 z^2+161828831232z-24461180928)\bigg\}.\\
 \end{aligned}
\end{equation}

Finally, it is illuminating to illustrate the asymptotic behaviors of the SDCs in the limit as 
$z$ approaches zero, which exhibit scalings as follows,
\begin{subequations}
	\begin{align}
	& d_{6,6}\left(c \rightarrow 0^{++}\right) =\frac{\pi ^2 \alpha_{s}^{4}}{108 z},
	\\
	& d_{3,3}\left(c \rightarrow 0^{++}\right) =\frac{\pi ^2 \alpha_{s}^{4}}{162 z},
	\\
	& d_{3,6}\left(c \rightarrow 0^{++}\right) =\frac{\pi ^2 \alpha_{s}^{4}}{54\sqrt{6} z},
	\\
	& d_{3,3}\left(c \rightarrow 1^{+-}\right) =\frac{13\pi ^2 \alpha_{s}^{4} z}{5832},
	\\
	& d_{3,3}\left(c \rightarrow 2^{++}\right) =\frac{4 \pi ^2 \alpha_{s}^{4}}{405 z}.
	\end{align}
	\label{limit-behavior}
\end{subequations}

\section{Phenomenology and discussion\label{phen-discu}}
\subsection{Choice of LDMEs}

Before making phenomenological predictions, it is essential to determine the values of the LDMEs that appear in 
the factorization formula \eqref{NRQCD:fac:t4c:fragmentation}. 
The most reliable first-principle approach to infer these matrix elements may be through lattice NRQCD simulation. Unfortunately, to date, no such lattice study has been conducted.
As an alternative, we turn to phenomenological methods to approximate these LDMEs. 

By applying the vacuum saturation approximation, the NRQCD LDMEs can be related to the wave functions at 
the origin~\cite{Feng:2023agq}
\begin{subequations}
	\begin{align}
& \left\langle {O}_{C_1,C_2}^{(0)}\right \rangle \approx 16\,{\psi_{C_1}(\mathbf{0})\psi_{C_2}^*(\mathbf{0})},
\\
& \left\langle {O}_{C_1,C_2}^{(1)}\right \rangle \approx 48\,{\psi_{C_1}(\mathbf{0})\psi_{C_2}^*(\mathbf{0})},
\\
& \left\langle {O}_{C_1,C_2}^{(2)}\right \rangle \approx 80\,{\psi_{C_1}(\mathbf{0})\psi_{C_2}^*(\mathbf{0})},
\end{align}
\label{LDMEs}
\end{subequations}
where $\psi(\mathbf{0})$ represents the four-body Schr\"odinger wave function at the origin. The color structure labels $C_1$ and $C_2$ can take on values of either $3$ or $6$, corresponding to the $\bar{\mathbf{3}}\otimes{\mathbf{3}}$ and ${\mathbf{6}}\otimes \bar {\mathbf{6}}$ diquark-antidiquark configuration. 
In phenomenology, the four-body Schr\"odinger wave functions for the fully-charmed tetraquarks have been extensively studied using various potential models. 
In this study, we enumerate the values of the LDMEs for various $S$-wave $T_{4c}$ states, including radially excited states up to the $3S$ level, derived from five 
distinct models~\cite{Lu:2020cns,Zhao:2020nwy,liu:2020eha,Yu:2022lak,Wang:2019rdo} in Table~\ref{LDME-value}. 
These values are sufficient for conducting phenomenological estimations. 
It is worth noting that the experimentally observed $X(6900)$ is more likely to be the $2S$ radial excited state~\cite{Lu:2020cns,Zhao:2020nwy,liu:2020eha,Yu:2022lak,Wang:2021kfv,Liu:2021rtn,Zhao:2020jvl}.
Upon examination of the table, it is observed that 
there is significant variation in the predictions of the four-body Schr\"odinger wave functions at the origin from different models,
albeit the mass spectra predicted by these models are closely aligned.
 
	
\begin{table}[!htbp]\normalsize
  \caption{ Numerical values of the LDMEs from different models, in unit of $\rm GeV^9$.}
  \label{LDME-value}
  \centering
  \setlength{\tabcolsep}{10pt}
  \renewcommand{\arraystretch}{2}
	\newcommand{\ldme}[3]{
		 \left\langle {O}_{#2,#3}^{(#1)}\right \rangle}
\begin{tabular}{cccccccccccccccc}
  \hline
  \hline
  \multirow{2}*{$nS$} & \multirow{2}*{$J^{PC}$} & \multirow{2}*{LDME} &\multicolumn{5}{c}{Models}\\
  \cline{4-8}
  &&&I~\cite{Lu:2020cns}&II~\cite{Zhao:2020nwy}&III~\cite{liu:2020eha}&IV~\cite{Yu:2022lak}&V~\cite{Wang:2019rdo}\\
  \hline
  \multirow{5}*{$1S$} & \multirow{3}*{$0^{++}$} & 
  $\ldme{0}{6}{6}$ & $0.0128$  & $2.50$  & $0.0027$ & $0.0173 $  & $0.00226$  \\
                   &           &
  $\ldme{0}{3}{6}$ & $0.0211$  & $3.65$  & $0.0033$ & $-0.0454 $ & $0.00215$  \\
                   &           &
  $\ldme{0}{3}{3}$ & $0.0347$  & $5.33$  & $0.0041$ & $0.119 $  & $0.00204$  \\
                   & $1^{+-}$  &
  $\ldme{1}{3}{3}$ & $0.0780$  & $12.6$  & $0.011$ & $0.0975 $  & $0.00876$  \\
                   & $2^{++}$  &
  $\ldme{2}{3}{3}$ & $0.072$   & $13.6$ & $0.012$  & $ 0.254$  & $0.0117$   \\
  \hline
  \multirow{5}*{$2S$} & \multirow{3}*{$0^{++}$} &
  $\ldme{0}{6}{6}$ & $0.0347 $ & $5.46$     & $0.0058$ & $0.0179 $  & $0.000545$ \\
                   &           &
  $\ldme{0}{3}{6}$ & $0.0538 $ & $10.6$     & $0.0067$  & $0.0386$   & $0.000890$ \\
                   &           &
  $\ldme{0}{3}{3}$ & $ 0.0832$ & $20.5$     & $0.0077$ & $0.0832$   & $0.00145$  \\
                   & $1^{+-}$  &
  $\ldme{1}{3}{3}$ & $0.1887 $ & $32.5$     & $0.021$   & $0.0648$   & $0.0173$   \\
                   & $2^{++}$  &
  $\ldme{2}{3}{3}$ & $0.1775 $ & $30.5$     & $0.026$  & $0.151$   & $0.0236$   \\
  \hline
  \multirow{5}*{$3S$} & \multirow{3}*{$0^{++}$} &
  $\ldme{0}{6}{6}$ & $-$       & $11.2$     & $-$       & $-$       & $-$        \\
                   &           &
  $\ldme{0}{3}{6}$ & $-$       & $22.5$     & $-$       & $-$       & $-$        \\
                   &           &
  $\ldme{0}{3}{3}$ & $-$       & $45.1$     & $-$       & $-$       & $-$        \\
                   & $1^{+-}$  &
  $\ldme{1}{3}{3}$ & $-$       & $39.3$     & $-$       & $-$       & $-$        \\
                   & $2^{++}$  &
  $\ldme{2}{3}{3}$ & $-$       & $42.5$     & $-$       & $-$       & $-$        \\

  \hline
  \hline
\end{tabular}
\end{table}

\subsection{$T_{4c}$ production at the \texttt{LHC}}
We can apply (\ref{t4c-cross section}) to calculate the $p_T$ spectrum of the $T_{4c}$ at the \texttt{LHC} with the center-of-mass (CM) energy $\sqrt{s}=13$ TeV. We take the charm quark mass $m_{c}=1.5$ $\rm GeV$, and the factorization scale 
$\mu=M_T$ with $M_T=\sqrt{p_T^2+M^2_{T_{4c}}}$ representing the transverse mass. To estimate the uncertainty from the scale, we slide $2 M_T \geq \mu\geq M_T/2$.
Moreover, a rapidity cut $5\geq y\geq -5$ is imposed. 
We utilize the \texttt{CT14 PDF}~\cite{Dulat:2015mca} sets for the proton PDF. 
Note that both partonic channels $gg\to c\bar{c}$ and $q \bar q\to c \bar c$ ($q$ = $u,d,s$) are considered in our computation.

It is widely accepted that in the high $p_T$ regime, fragmentation production becomes the dominant mechanism. 
Notably, an observation suggests that the production rate of the $T_{4c}$ via the fragmentation
mechanism may begin to predominate at $p_T > 20$ GeV, as indicated in \cite{Feng:2023agq}.

\begin{figure}[H]
	\captionsetup{list=off,format=cont}
	\captionsetup[subfigure]{format=plain}
  {\includegraphics[width=0.45\linewidth]{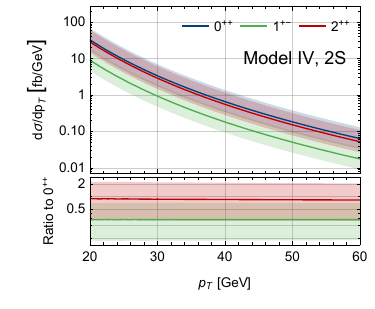}}\hfill
  {\includegraphics[width=0.45\linewidth]{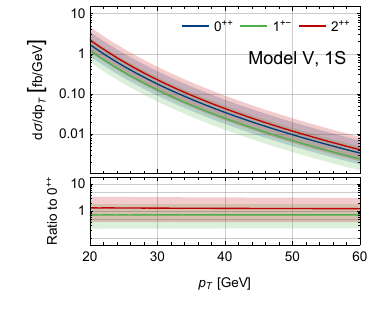}}\\
  {\includegraphics[width=0.45\linewidth]{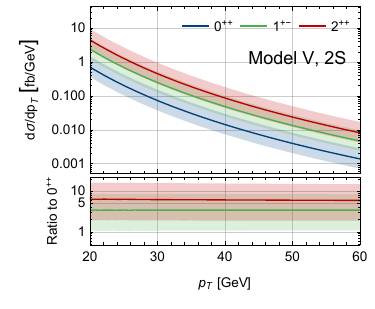}}\hfill
  \caption{The $p_T$ spectra of various $S$-wave $T_{4c}$ states at the \texttt{LHC} with $\sqrt{s}=13\ \mathrm{TeV}$ predicted using five distinct potential models.  The blue, green and red curves correspond to the differential cross sections for the $0^{++}$, $1^{+-}$ and $2^{++}$ tetraqaurks, respectively. 
  The lower inset in each plot displays the ratios of the differential cross sections for the $1^{+-}$ and $2^{++}$ tetraquarks to that of the $0^{++}$.  }
  \label{fig:pt_dist}
\end{figure}

In figure~\ref{fig:pt_dist}, we present the $p_T$ distributions for the various
$T_{4c}$ states at the \texttt{LHC}, focusing on $p_T > 20$ GeV.
Upon examining the figure, it is observed that the cross section diminishes markedly with increasing transverse momentum $p_T$. 
Additionally, the differential cross sections for the distinct $T_{4c}$ states exhibit similar $p_T$-dependent trends.

\begin{figure}[H]
  \centering
  {\includegraphics[width=0.45\linewidth]{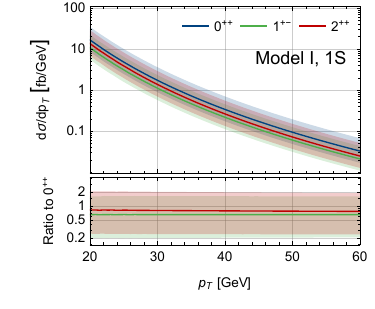}}\hfill
  {\includegraphics[width=0.45\linewidth]{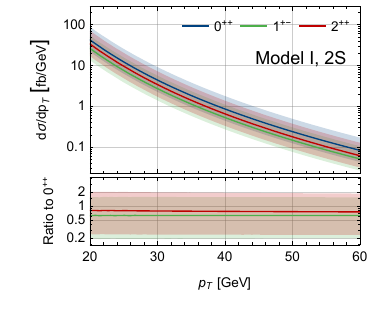}}\\
  {\includegraphics[width=0.45\linewidth]{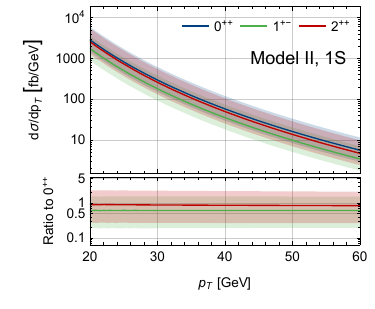}}\hfill
  {\includegraphics[width=0.45\linewidth]{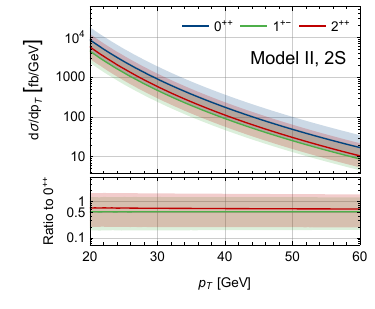}}\\
  {\includegraphics[width=0.45\linewidth]
  {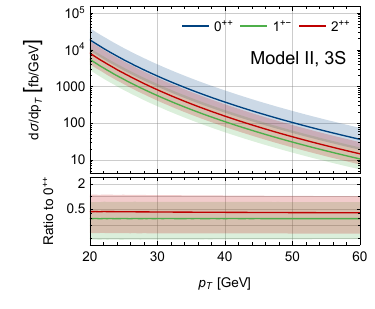}}\hfill
  {\includegraphics[width=0.45\linewidth]{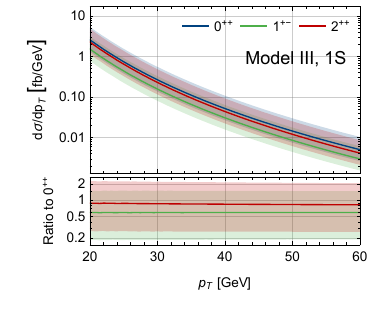}}\\
  {\includegraphics[width=0.45\linewidth]{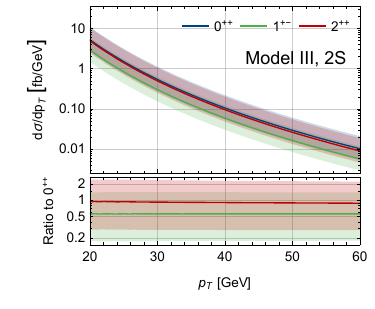}}\hfill
  {\includegraphics[width=0.45\linewidth]{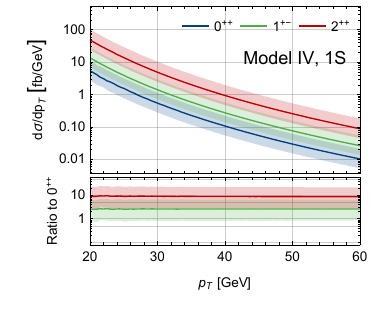}}
\end{figure}

Now, we proceed to predict the integrated cross sections. 
We can implement a kinematic cut at $p_T > 20$ GeV to predict the
integrated cross section over the transverse momentum range.
The cross sections derived from various potential models are presented in Table~\ref{cross-prediction-LHC}.
To facilitate a comparison of their relative magnitude, we have listed the contributions from the partonic channels $gg\to c\bar{c}$ and  
$q\bar{q}\to c\bar{c}$ separately.
An analysis of the data presented in the table reveals several key observations. 

	\begin{table}[!htbp]\footnotesize
		\caption{The integrated cross sections for $T_{4c}$ fragmentation production at the \texttt{LHC}, calculated from the partonic channels 
		$gg \to c\bar{c}$ and $q \bar{q} \to c\bar{c}$ and denoted as $gg$ and $q\bar{q}$ for brevity,  are presented in unit of fb 
		for $p_T > 20$ GeV. 
		To ensure all data fits within the table, we have rescaled the cross sections by applying an appropriate multiplication factor. Additionally, 
		we have taken the central value of the cross section to the mid-value of the factorization scale variation.
		}
		\label{cross-prediction-LHC}
		\centering
		\setlength{\tabcolsep}{2.5pt}
		\renewcommand{\arraystretch}{2}
		\begin{tabular}{ccccccccccccc}
			\hline
			\hline
			\multirow{2}{*}{Model}& \multirow{2}{*}{Channel} & \multicolumn{3}{c}{$1S$} & \multicolumn{3}{c}{$2S$} & \multicolumn{3}{c}{$3S$} \\
			\cline{3-11}
			&                      & $0^{++}$        & $1^{+-}$      & $2^{++}$        & $0^{++}$      & $1^{+-}$      & $2^{++}$ & $0^{++}$ & $1^{+-}$ & $2^{++}$ \\
			\hline
			\multirow{2}{*}{I~\cite{Lu:2020cns}}
			& $gg(\times 0.1)$
			& $9.3\pm5.7$          & $6.0\pm3.6$     & $7.6\pm4.7$   & $24\pm14$       & $15\pm9$      & $19\pm12$     & $-$      & $-$       & $-$                  \\
			& $q\bar{q}(\times10)$   & $10\pm 6$       & $6.6\pm3.7$
			& $9.4\pm5.5$          & $26\pm15$       & $16\pm9$      & $23\pm13$       & $-$            & $-$            & $-$                                        \\
			\hline
			\multirow{2}{*}{II~\cite{Zhao:2020nwy}}
			& $gg(\times 10^{-3})$
			& $16\pm10$            & $9.7\pm5.9$        & $14\pm9$     & $48\pm29$     &$25\pm15$     & $32\pm20$    & $100\pm60$     &$30\pm18$  &$45\pm28$                 \\
			& $q\bar{q}(\times 0.1)$
			& $18\pm10$            & $11\pm6$        & $18\pm10$       &$54\pm31$      &$27\pm16$      &$40\pm23$    & $ 120\pm70$      &$33\pm19$     & $55\pm32$            \\
			\hline
			\multirow{2}{*}{III~\cite{liu:2020eha}}
			& $gg$
			& $14\pm9$             & $8.5\pm5.1$     & $13\pm8$      & $29\pm18$       & $16\pm10$     & $27\pm17$     & $-$       & $-$       & $-$                 \\
			& $q\bar{q}(\times10^2)$
			& $16\pm9$             & $9.3\pm5.3$     & $16\pm9$      & $32\pm18$       & $18\pm10$     & $34\pm20$     & $-$        & $-$        & $-$                  \\
			\hline
			\multirow{2}{*}{IV~\cite{Yu:2022lak}}
			& $gg(\times 0.1)$
			& $2.9\pm1.8$            & $7.5\pm4.5$       & $27\pm17$   & $18\pm11$     & $5\pm3$     & $16\pm10$   & $-$      & $-$        & $-$                   \\
			& $q\bar{q}(\times10)$
			& $3.3\pm1.9$        & $8.2\pm4.7$   & $33\pm19$   & $20\pm11$     & $5.5\pm3.1$ & $20\pm11$   & $-$      & $-$       & $-$                  \\
			\hline
			\multirow{2}{*}{V~\cite{Wang:2019rdo}}
			& $gg$
			& $9.5\pm5.7$          & $6.7\pm4.1$     & $12\pm8$      & $3.9\pm2.4$     & $13\pm8$      & $25\pm15$     & $-$        & $-$       & $-$                   \\
			& $q\bar{q}(\times10^2)$
			& $11\pm6$        & $7.4\pm4.2$ & $15\pm9$ & $4.4\pm2.5$ & $15\pm8$ & $31\pm18$ & $-$       & $-$        & $-$               \\
			\hline
			\hline
		\end{tabular}
		
	\end{table}

Firstly, it is observed that the predicted cross sections for $T_{4c}$ production exhibit a strong dependence on phenomenological models. Notably, 
the predictions from Model II are significantly larger than those from other models, due to the substantially larger LDMEs.
In most cases, the predictions from Model I and IV are closely aligned and approximately one order of magnitude larger than those from Model III and V. 
Concretely, the cross sections for $T_{4c}$ production from Model III and V are found to be in the range of $5$ to $30$ fb, whereas
Model I and IV predict the cross sections mostly in the range of $100$ to $300$ fb. The predictions from Model II extend even to $10$ pb.  
There is no doubt that the primary theoretical uncertainties originate from the imprecision of the LDMEs.  Consequently, it is 
essential to accurately determine the LDMEs to enable precise theoretical predictions.  

Secondly, it is evident that the contributions via the partonic channel $gg\to c\bar{c}$
are significantly predominant over those from $q\bar{q}\to c\bar{c}$, making the latter contributions negligible for practical purposes. 
This predominance can be attributed to the fact that the gluon PDF is considerably larger than light quark PDFs in the kinematic regime of the considered processes.

Finally, it is insightful to compare the relative magnitudes of the $T_{4c}$ production rates via charm quark fragmentation and via gluon fragmentation at the \texttt{LHC}. 
Our analysis indicates that, assuming identical LDMEs,
 the cross section for $T_{4c}$ production from charm quark fragmentation is significantly smaller than that from gluon fragmentation. 
 The cross section for the latter process has been calculated in Ref.~\cite{Feng:2020riv}.
This discrepancy can be attributed to two primary factors. 
Firstly, the partonic cross section for $gg\to c\bar{c}$, as given in equation (\ref{eq-partonic-gg}), is considerably smaller than that for $gg\to gg$ across most
$p_T$ regimes. 
An additional factor to consider is that the magnitude of the fragmentation function for $c\to T_{4c}$ is comparatively smaller than that for $g\to T_{4c}$.

\subsection{$T_{4c}$ production at $ep$ colliders}

As another application of the fragmentation function,  we 
predict the integrated cross sections for $T_{4c}$ photoproduction at the  \texttt{HERA}, \texttt{EIC}, and \texttt{EicC}.
At lowest order in $\alpha_s$, $T_{4c}$ can be produced through either the partonic process $\gamma q\to gq$ followed by gluon fragmentation
into $T_{4c}$, or the process $\gamma g\to c\bar{c}$ followed by charm quark fragmentation into $T_{4c}$.

We employ equation (\ref{t4c-cross section-ep}) to evaluate the cross section, taking 
the electron mass $m_{e}=0.51\times 10^{-3}$ GeV, and the fine structure constant $\alpha =\frac{1}{137}$.  
Furthermore, we impose transverse momentum constraint $p_T> 20$ GeV, and cut on the elasticity parameter:  $0.9>z>0.3$ for \texttt{HERA}, $0.9>z>0.05$ for \texttt{EIC}, and $0.7>z>0.05$ for \texttt{EicC},
to eliminate contributions from both diffractive and resolved-photon processes. 
Additionally, to ensure photoproduction-type events, we apply cuts on the photon virtuality, similar to those used in the photoproduction of charmonia
~\cite{Flore:2020jau}: $Q^2_{max}=2.5\ \mathrm{GeV}^2$ for \texttt{HERA}, and $Q^2_{max}=1\ \mathrm{GeV}^2$ for \texttt{EIC} and \texttt{EicC}.
For other cut-irrelevant parameters, we adhere to the same choices as in the $pp$ collision.

The integrated cross sections for $T_{4c}$ fragmentation production at the \texttt{HERA} and \texttt{EIC} are listed
in  Table~\ref{cross-prediction-ep}.~\footnote{Since the CM energy of the \texttt{EicC} is around $20$ GeV, which is less than the 
$p_T$ cut,  theoretical predictions for this collider are not included in Table~\ref{cross-prediction-ep}.}
Upon examining the table,  we have several observations. 
Firstly,  the cross sections for $T_{4c}$ production through the partonic process $\gamma g\to c\bar{c}$ 
are substantially smaller than those obtained through  the partonic process $\gamma q\to gq$. 
This is attributed to the fact that the cross section of the partonic process $\gamma q\to gq$ is larger than that of $\gamma g\to c\bar{c}$, 
and additionally the gluon fragmentation function is considerably greater than the charm quark fragmentation function.
Secondly, the cross sections for $T_{4c}$ production at the \texttt{EIC} are approximately one order of magnitude smaller than those
for production at the \texttt{HERA}. This is primary due to the difference in the CM energy.
Thirdly, it is noted
that the cross sections for $T_{4c}$ production at $ep$ colliders are significantly smaller than those at the \texttt{LHC}.
Note, except for the predictions from Model II, the cross sections for $T_{4c}$ production at $ep$ colliders are found to be less than $1$ fb. 
Finally, similar to what is observed at the \texttt{LHC}, there are larger uncertainties in the cross sections.
The significant variance in the values of LDMEs is the primary source of theoretical uncertainty in the cross section, substantially eclipsing the uncertainty associated with the scale.

Since the differential cross sections increase significantly as the transverse momentum $p_T$ decreases,  
it is expected that a substantial fraction of the cross section will be concentrated in the moderate $p_T$ range.
 However, in this range, the fragmentation approximation is no longer applicable, 
 necessitating the use of fixed-order perturbative calculations.
The $1^{+-}$ $T_{4c}$ is dominantly produced through the process $\gamma g\to T_{4c}g$, 
and considerable cross sections at the \texttt{HERA} and \texttt{EIC} have been reported in reference~\cite{Feng:2023ghc}. 
Due to $C$ parity conservation, $T_{4c}$ states with $0^{++}$ and $2^{++}$ only begin to be produced through 
$2\to 3$ partonic processes, such as $\gamma g\to T_{4c}g g$, $\gamma g\to T_{4c} c\bar{c}$, and $\gamma q\to T_{4c} g q$.
This makes the computation of these cross sections much more complex.   
Nevertheless, using the formulas presented in this work, we can approximately 
estimate the cross sections for $0^{++}$ and $2^{++}$ tetraquark production. 
Although the fragmentation mechanism is not valid at low $p_T$, 
it is reasonable to expect the cross sections predicted by the fragmentation mechanism are order-of-magnitude
consistent with those obtained by fixed-order perturbative calculations.   
By taking $p_T> 6$ GeV, we present the summed contributions of both fragmentation channels to the cross sections in Table~\ref{cross-prediction-ep-6gev}.
Upon inspection of the table,  we find the cross sections at the \texttt{HERA} and \texttt{EIC} are moderate.
 For instance,
the cross sections for certain $T_{4c}$ production, as predicted by Model I, can reach $10$ fb at the \texttt{HERA} 
and $5$ fb at the \texttt{EIC}. In contrast, the cross sections for $T_{4c}$ production at the \texttt{EicC} are small. 
Given the integrated luminosity of $468\, {\rm pb}^{-1}$ at the \texttt{HERA} and $100\, {\rm fb}^{-1}$ at the \texttt{EIC} for 
one year of data collection, it seems that detection of these fully heavy tetraquarks at $ep$ colliders remains somewhat challenging.

		\begin{table}[!htbp]\footnotesize
		\caption{The $p_{T}$-integrated cross sections for $T_{4c}$ photoproduction at the \texttt{HERA} and \texttt{EIC} are presented, with units in fb, 
			The CM energies of 
			\texttt{HERA} and \texttt{EIC} are chosen to be $319$ GeV and $140.7$ GeV, respectively. 
			$T_{4c}$ can be produced through either the partonic process $\gamma q\to gq$, followed by $g\to T_{4c}$, or the process $\gamma g\to c\bar{c}$,
			followed by $c\to T_{4c}$. For brevity, we denote the former process $\gamma q$, 
			and the latter process $\gamma g$. 
            Note $p_T$ is limited to be larger than $20\ \mathrm{GeV}$ to ensure the validity of the fragmentation mechanism. 
		}
		\label{cross-prediction-ep}
		\centering
		\setlength{\tabcolsep}{1.3pt}
		\renewcommand{\arraystretch}{2}
		\begin{tabular}{ccccccccccccccccc}
		\hline
		\hline
		&\multirow{2}*{$nS$}&\multirow{2}*{ $J^{PC}$}
		&\multicolumn{2}{c}{I~\cite{Lu:2020cns}} & \multicolumn{2}{c}
		{II~\cite{Zhao:2020nwy}} & \multicolumn{2}{c}
		{III~\cite{liu:2020eha}} & \multicolumn{2}{c}
		{IV~\cite{Yu:2022lak}} & \multicolumn{2}{c}
		{V~\cite{Wang:2019rdo}} \\
		\cline{4-13}
		&                     &
		& $\gamma q(\times10^2)$       & $\gamma g(\times10^3)$ 
		& $\gamma q$   & $\gamma g$     & $\gamma q(\times10^3)$ & $\gamma g(\times10^4)$ & $\gamma q(\times10^2)$ & $\gamma g(\times10^4)$ & $\gamma q(\times10^3)$ & $\gamma g(\times10^5)$ \\
		\hline
		\multirow{9}*{$\rm HERA$}
		& \multirow{3}*{$1S$} & $0^{++}$
		& $1.5$               & $1.5$     & $2.6$ & $0.25$  
		& $2.3$         & $2.2$         & $0.48$         & $4.6$         & $1.5$         & $15$          \\
		&                     & $1^{+-} $
		& $-$               & $0.91$        & $-$         & $0.15$        
		& $-$  & $1.3$  & $-$         & $11$         & $-$          & $10$ 
		\\
		&                     & $2^{++} $
		& $2.5$               & $0.89$     & $4.7$  & $0.17$ 
		& $4.2$         & $1.5$         & $8.8$          & $31$          & $4.1$         & $14$          \\
        \cline{2-13}
		& \multirow{3}*{$2S$} & $0^{++}$
		& $3.8$               & $3.7$        & $7.8$   & $0.75$     & $4.7$          & $4.5$         & $2.9$         & $28$          & $0.64$         & $6.1$         \\
		&                     & $1^{+-} $
		& $-$               & $2.2$        & $-$         & $0.38$        
		& $-$  & $2.5$  & $-$         & $7.6$         & $-$          & $20$ 
		\\
		&                     & $2^{++}$
		& $6.2$                & $2.2$         & $11$   &  $0.38$  
		& $9.0$          & $3.2$         & $5.2$          & $19$          & $8$          & $29$          \\
        \cline{2-13}
		& \multirow{3}*{$3S$} & $0^{++}$
		& $ -$                 & $- $            & $17$   &  $1.6$    & $-$ & $-$ & $-$ &$-$&  $-$  &$-$                    
		\\
		&                     & $1^{+-} $
		& $-$               & $-$        & $-$         & $0.46$        
		& $-$  & $-$  & $-$         & $-$         & $-$          & $-$ \\
		&                     & $2^{++}$
		& $- $                 & $- $            & $15$   &  $0.52$    & $-$ & $-$ & $-$ &$-$&  $-$  &$-$                                \\
		\hline
		\multirow{9}*{$\rm EIC$}
		& \multirow{3}*{$1S$} & $0^{++}$
		& $0.21$              & $0.13$       & $0.36$ & $0.021$ & $0.33$        & $0.19$        & $0.067$        & $0.39$        & $0.22$        & $1.3$         \\
		&                     & $1^{+-} $
		& $-$               & $0.076$        & $-$         & $0.012$        
		& $-$  & $0.11$  & $-$         & $0.95$         & $-$          & $0.85$ \\
		&                     & $2^{++} $
		& $0.37$              & $0.069$        & $0.70$ & $0.013$ & $0.62$         & $0.12$        & $1.3$         & $2.4$         & $0.60$         & $1.1$         \\
        \cline{2-13}
		& \multirow{3}*{$2S$} & $0^{++}$
		& $0.54$               & $0.32$              & $1.1$   &  $0.065$  & $0.66$         & $0.39$        & $0.41$        & $2.4$         & $0.089$        & $0.53$        \\
		&                     & $1^{+-} $
		& $-$               & $0.18$        & $-$         & $0.032$        
		& $-$  & $0.21$  & $-$         & $0.63$         & $-$          & $1.7$ \\
		&                     & $2^{++}$
		& $0.91$               & $0.17$              & $1.6$   & $0.029$  & $1.3$         & $0.25$        & $0.78$         & $1.5$         & $1.2$         & $2.3$         \\
        \cline{2-13}
		& \multirow{3}*{$3S$} & $0^{++}$
		& $- $                 & $ -$               & $2.4$  &$0.14$
		&  $-$ & $-$ & $-$ &$-$&  $-$  &$-$                                                                                                            \\
		&                     & $1^{+-} $
		& $-$               & $-$        & $-$         & $0.038$        
		& $-$  & $-$  & $-$         & $-$         & $-$          & $-$ 
		\\
		&                     & $2^{++}$
		& $- $                 & $- $               & $2.2$  &$0.041$
		&  $-$ & $-$ & $-$ &$-$&  $-$  &$-$ \\
		\hline
		\hline
		\end{tabular}
		\end{table}

		\begin{table}[!htbp]\normalsize
		\caption{The $p_{T}$-integrated cross sections for $T_{4c}$ photoproduction at the \texttt{HERA}, \texttt{EIC} and \texttt{EicC} are presented, with units in fb.
	The CM energies of 
	\texttt{HERA}, \texttt{EIC}, and \texttt{EicC} are chosen to be $319$ GeV, $140.7$ GeV and $20$ GeV, respectively. 
        $p_T$ is set to be larger than $6\ \mathrm{GeV}$ to estimate the production cross sections in the moderate $p_T$ range. 
		}
		\label{cross-prediction-ep-6gev}
		\centering
		\setlength{\tabcolsep}{7pt}
		\renewcommand{\arraystretch}{2}
		\begin{tabular}{ccccccccccccccccc}
		\hline
		\hline
		&\multirow{1}*{$nS$}&\multirow{1}*{ $J^{PC}$}
		&\multirow{1}*{I~\cite{Lu:2020cns}} & 
	\multirow{1}*{II~\cite{Zhao:2020nwy}}
		& \multirow{1}*{III~\cite{liu:2020eha}} & 
		\multirow{1}*{ IV~\cite{Yu:2022lak}} &
		\multirow{1}*{ V~\cite{Wang:2019rdo}} \\
		\cline{4-8}
	    \hline
		\multirow{6}*{$\rm HERA$}
		& \multirow{2}*{$1S$} & $0^{++}$
		& $2.7$               & $470$           & $0.42$ & $0.88$ & $0.28$  \\
		&                     & $2^{++} $
		& $3.8$               & $710$            & $0.63$ & $13.3$ & $0.61$  \\
		& \multirow{2}*{$2S$} & $0^{++}$
		& $7.0$                & $1400$         & $0.86$    & $5.4$   & $0.12$ \\
		&                     & $2^{++}$
		& $9.3$                & $1600$           & $1.4$    & $7.9$  & $1.2$   \\
		& \multirow{2}*{$3S$} & $0^{++}$
		& $- $                 & $ 3100$             & $-$    &  $-$  &  $ -$     \\
		&                     & $2^{++}$
		& $- $                & $ 2200$             & $-$    & $-$   &   $ -$   \\
		\hline
		\multirow{6}*{$\rm EIC$}
		& \multirow{2}*{$1S$} & $0^{++}$
	    & $1.3$               & $220$           & $0.19$ & $0.40$ & $0.13$  \\
	    &                     & $2^{++} $
	    & $1.7$               & $320$            & $0.28$ & $6.0$ & $0.28$  \\
   	    & \multirow{2}*{$2S$} & $0^{++}$
    	& $3.2$                & $650$         & $0.39$    & $2.5$   & $0.053$ \\
    	&                     & $2^{++}$
    	& $4.2$                & $720$           & $0.62$    & $3.6$  & $0.56$   \\
    	& \multirow{2}*{$3S$} & $0^{++}$
    	& $-$                 & $1400 $             & $-$    &  $-$  &  $ -$     \\
    	&                     & $2^{++}$
    	& $- $                & $ 1000$             & $-$    & $-$   &   $ -$   \\
    	\hline
	   	\multirow{6}*{$\rm EicC$}
	    & \multirow{2}*{$1S$} & $0^{++}$
	    & $4.8\times 10^{-5}$    & $8.1\times 10^{-3}$ 
           & $7.4\times 10^{-6}$ & $1.5\times 10^{-5}$ & $4.8\times 10^{-6}$  \\
	    &                     & $2^{++} $
	    & $8.0\times 10^{-5}$    & $1.5\times 10^{-2}$            & $1.3\times 10^{-5}$ & $2.8\times 10^{-4}$ & $1.3\times 10^{-5}$  \\
	    & \multirow{2}*{$2S$} & $0^{++}$
	    & $1.2\times 10^{-4}$     & $2.5\times 10^{-2}$         & $1.5\times 10^{-5}$    & $9.3\times 10^{-5}$   & $2.0\times 10^{-6}$ \\
	    &                     & $2^{++}$
	    & $2.0\times 10^{-4}$      & $3.4\times 10^{-2}$           & $2.9\times 10^{-5}$    & $1.7\times 10^{-4}$  & $2.6\times 10^{-5}$   \\
	    & \multirow{2}*{$3S$} & $0^{++}$
	    & $-$                 & $5.3\times 10^{-2} $             & $-$    &  $-$  &  $ -$     \\
	    &                     & $2^{++}$
	    & $-$                & $ 4.7\times 10^{-2}$             & $-$    & $-$   &   $- $   \\                    
		\hline
		\hline
		\end{tabular}
		\end{table}
		

\newpage
\section{Summary\label{summary}}

In this study, we calculate the fragmentation function of a charm quark into an $S$-wave fully-charmed tetraquark $T_{4c}$, 
using the framework of NRQCD.  
The SDCs, which are universal, are determined at the lowest order in $\alpha_s$ and $v$. 
The nonperturbative LDMEs can be approximated by the $T_{4c}$ four-body wave 
functions at the origin, which have been evaluated using phenomenological potential models in literature.
The fragmentation function derived in this work can be utilized to predict the $T_{4c}$ production cross sections
at various particle colliders. 

As a first application, we compute the cross sections for $T_{4c}$ production at the \texttt{LHC}.
To investigate model dependence, we employ five phenomenological potential models to determine the LDMEs.
Both the differential distribution over the transverse momentum of $T_{4c}$ and the integrated cross sections are presented. 
Despite significant uncertainties arising from the LDMEs,  the calculated cross sections are substantial, 
suggesting that a considerable number of events could be generated at the \texttt{LHC}.

Furthermore, we have estimated the cross sections for the photoproduction of the  $T_{4c}$ at the \texttt{HERA}, \texttt{EIC}, and \texttt{EicC}. 
These cross sections from most models are found to be moderate at the \texttt{HERA} and \texttt{EIC}, while be quite small at the \texttt{EicC}.
Given the integrated luminosities at these colliders,
 the detection of fully-charmed tetraquarks in $ep$ colliders is somewhat challenging.

\section*{Acknowledgments}
We are grateful to Ming-Sheng Liu, Qi-Fang Lü, and
Jiaxing Zhao, Guo-Liang Yu and Lu Meng for providing us with the values of tetraquark
wave functions at the origin from various potential models.
The work of X.-W. B.,  C.-M. Gan, and W.-L. S. is supported by the
National Natural Science Foundation of China under Grants 
No. 12375079 and No. 11975187, and the Natural Science
Foundation of ChongQing under Grant No. CSTB2023
NSCQ-MSX0132. The work of F. F. is supported by the National Natural Science Foundation of China under Grant
No. 12275353. The work of Y.-S.~H. is supported by the DOE grants DE-FG02-91ER40684 and DE-AC02-06CH11357.

\end{document}